\documentclass[12pt,titlepage]{article}%
\pdfoutput=1 
\usepackage{amsfonts}
\usepackage{amsmath}
\usepackage{amssymb}
\usepackage{graphicx}%
\providecommand{\U}[1]{\protect\rule{.1in}{.1in}}

\begin{document}

\title{LIGO-India: A Decadal Assessment on Its Scope, Relevance, Progress, and Future}
\author{C. S. Unnikrishnan\thanks{With reference to my article `IndIGO and LIGO-India:
Scope and Plans for Gravitational Wave Research and Precision Metrology in
India', Int. Jl. Mod. Physics D, Vol. 22, 1341010 (2013).\ E-mail address:
unni@tifr.res.in}\\Tata Institute of Fundamental Research, Mumbai 400005, India\thanks{Presently
at School of Quantum Technology, DIAT (DRDO), Girinagar, Pune 411025. E-mail
address: unni@diat.ac.in}}
\date{LIGO-P2200353-v3}
\maketitle

\begin{abstract}
The LIGO-India project to build and operate an advanced LIGO (aLIGO)
gravitational wave (GW) detector in India in collaboration with LIGO-USA was
considered and initiated as an Indian national megascience project in 2011.
The project relied on the advantage of a head start and cost saving due to the
ready availability of the full aLIGO interferometer components from LIGO-USA,
approved by the National Science Foundation. Procedural formalities and site
selection efforts progressed since then and the provisional approval for the
Indian national project was obtained in 2016, immediately following the first
direct detection of gravitational waves with the aLIGO detectors. I had
discussed the scope and the science case of the project, as well as the plans
for its realization by 2022, in several talks and in a paper (Int. Jl. Mod.
Physics D, Vol. 22, 1341010, 2013). With KAGRA GW detector in Japan being
tuned to be part of the GW detector network, it is now the occasion to assess
the progress of LIGO-India project, and evaluate its relevance and scope for
gravitational wave science and astronomy. Various key factors like
human-power, management, funding, schedule etc., in the implementation of the
project are reassessed in the backdrop of the evolution of the global
GW\ detector sensitivity. In what I consider as a realistic estimate, it will
take more than a decade, beyond 2032, to commission the detector even with a
fraction of the projected design sensitivity. I estimate that the budget for
implementation will be more than doubled, to about Rs. 35 billion (%
$>$%
3500 crores, \$430 million). The detrimental consequences for the project are
discussed, from my personal point of view, as the coordinator of the
experimental and technical aspects in the 2011 LIGO-India proposal. My decadal
assessment points to the eroded scientific relevance of the LIGO-India
detector in the GW\ detector network\ of the next decade. However, a revamped
action plan with urgency and the right leadership can make LIGO-India a late
but significant success for multi-messenger astronomy for several years after
2032, because of its design similitude to the operational aLIGO detectors. For
achieving this, it is imperative that the LIGO-India detector is replanned and
launched in the post-O5 upgraded A$^{\#}$ version, similar to the projected
LIGO-USA detectors.

\medskip

Keywords: Gravitational wave detectors, LIGO detectors, LIGO-India project, GW astronomy

\end{abstract}

\section{Preamble}

The second generation interferometric detectors of gravitational waves were
the result of major upgrades in the design and implementation of the LIGO
detectors in the USA\ and the Virgo detector in Europe, during 2009-2015. The
enthusiastic surge of interest in\ the anticipated detection of gravitational
waves (GW) with advanced LIGO (aLIGO) detectors and the advanced Virgo
detector led to the LIGO-India project proposal in 2011, to build and operate
a third advanced LIGO detector in India. The proposal was the realization of
`now, or never' about gravitational wave science and astronomy, among the
physicists who initiated the proposal \cite{Proposal-DCC,Unni-LI,Current-LI}.
The discussions in small meetings had given birth to the \emph{Indigo
consortium} of the GW\ research community in India, which gained visibility of
their seriousness very soon. We were encouraged, trusted, and supported by the
GW community globally, and in particular by the GWIC (GW\ International
Committee) and the scientists from the LIGO laboratory. The LIGO project came
to India after an active discussion during 2010-11 on the possibility of
LIGO-Australia, in which the participation of India was proposed to be a
collaborating partner. The idea of LIGO-Australia was that the interferometer
components that were meant for a third LIGO detector at the same site as the
US-Hanford detector should be used to build the third detector at a site in
Australia, which would enable the accurate localization of GW\ events due to
the availability of a third independent detector at a large baseline distance
\cite{LIGO-Aus}. This was a major decision that prioritized and emphasized GW
astronomy, rather than just the detection of the gravitational waves. The
entire infrastructure and the operation of the detector in the network were
the responsibility of the host. LIGO-Australia finally turned out to be not
feasible,\ mainly due to the funding constraints. Then the Indigo consortium
and the LIGO-laboratory explored the possibility of a LIGO detector in India.
A high level committee that advised the National Planning Commission of the
Government of India on large and long term national projects in astronomy
recommended the LIGO-India project, after multiple presentations and
evaluation, fully realizing its importance and the unprecedented advantage of
obtaining the full interferometer detector components from the LIGO laboratory
in the USA.

The main infrastructural elements for building the detector are a suitable
site, laboratory buildings and the enormous ultra-high vacuum (UHV) enclosures
of the 4 km x 4 km Michelson interferometer, for which participating Indian
institutions had the full responsibility. Site selection efforts started
immediately by the Indigo consortium members, even before any assurance of
funding, with timely support from the Inter-University Centre for Astronomy
and Astrophysics (IUCAA), Pune. The identification of a few very good sites,
isolated well from common noise sources, followed after the elimination of
many which were considered, based on several visits and preliminary
measurements of seismic noise. When IUCAA, Pune, and the two key
technologically highly endowed institutes under the Department of Atomic
Energy (DAE) -- The Institute for Plasma Research (IPR), Gandhinagar, and the
Raja Ramanna Centre for Advanced Technology (RRCAT), Indore -- agreed to take
key responsibilities for the projects, things came together as a feasible
project. Speedy and cautious response and support from the NSF, USA, in the
form of visits of key persons and reviews of the proposal by special
committees gave a concrete form to the LIGO-India project. Four senior level
visits from the LIGO-Laboratory to the LIGO-India lead-institutions for
technical assessment and discussions were followed by three in depth reviews
by a NSF panel. All this culminated in a review and the following resolution
by the National Science Board, USA, in August 2012: \textquotedblleft
Resolved, that the National Science Board authorize the Deputy Director at her
discretion to approve the proposed Advanced LIGO Project change in scope,
enabling plans for the relocation of an advanced detector to
India\textquotedblright.

The responsibility for the data handling, computing as well as the important
task of building and coordinating the user community for GW astronomy was also
taken by IUCAA. With the technical aspects of UHV, seismic isolation and
suspensions, optical assembly, control systems etc. being handled by IPR,
Gandhinagar and RRCAT, Indore, the LIGO-India project had a defined
institutional structure. Regular visits by senior scientists of the LIGO
laboratory and planning meetings helped in charting a road map and organizing
a strong user community within a short period. Hence, three\textbf{
}significant developments happened by the time the first detection of
gravitational waves was announced in early 2016; 1) the Indigo consortium and
the interest in GW astronomy had grown considerably, ten-fold during
2011-2016, 2) a large number of Indian researchers were already members of the
global LIGO Scientific Collaboration (LSC) and part of the discovery
\cite{Discovery2016,Unni-bonanza}, 3) a suitable site for the LIGO-India
detector was identified in the state of Maharashtra and finalized (Aundha in
Hingoli district, Latitude 19%
${{}^\circ}$
36' 50\textquotedblright\ N, Longitude 77%
${{}^\circ}$
01' 50\textquotedblright\ E, about 175 hectares). The prompt announcement by
the Prime Minister of India of the formal support and approval for the project
then provided the boost and fuel to go ahead. The funding responsibility,
which was estimated then as about Rs. 13 billion (Rs. 1300 Crores, \$270
million), was to be shared between the DAE and the Department of Science \&
Technology (DST). The contribution from NSF and the LIGO laboratory was the
entire set of aLIGO detector components (except the vacuum infrastructure)
including the stabilised laser, to be given in kind. The notional cost of the
contribution of the LIGO laboratory and other international partners in the
LIGO Scientific Collaboration was about \$ 130 million.

\section{The LIGO-India Project: The Science Case}

The strong science case for the LIGO-India (LI) project had been discussed on
many occasions and in publications \cite{Proposal-DCC, Unni-LI,Fair-Sathya}.
The project proposal itself is available at the website of LIGO-India and the
Indigo consortium (gw-indigo.org). An early compilation, leaning on the
contributions from many from the LIGO Scientific Collaboration as well as the
Indigo Consortium, is in my 2013 overview paper, based on a talk in the Astrod
symposium (2012) held in Bengaluru \cite{Unni-LI}. The most important aspect
of LIGO-India is that it is the only planned detector with the design
sensitivity matching the aLIGO detectors, and of identical design and
technological characteristics. Therefore, if implemented, it provides assured
detections with excellent source localization on a single integrated platform.
The other aspects of the additional detector in the network are the
significantly improved network duty cycle, improved sensitivity, and larger
sky coverage. The upgraded version of the European Virgo detector was already
getting commissioned for joint observations with the two LIGO detectors, and
the Japanese detector KAGRA \cite{Kagra-1} was nearing completion, when the
LIGO-India project was approved in 2016. The hope was that a network of all
these five detectors would become operational by 2022-23.

The sensitivity of the individual detectors can be compared when stated in
terms of the equivalent range for the detection of the merger of binary
neutron stars (BNS). The design sensitivity of the aLIGO detectors is about
170 Mpc. With another planned upgrade with a frequency-dependent squeezed
light technology and improved mirror coating, called LIGO A+, the mature
sensitivity is projected to reach beyond 300 Mpc. Advanced LIGO started its
operation (O1) in 2015 with a range above 60 Mpc, and progressively reached a
sensitivity of 130 Mpc. The fourth observing cycle (O4) will start in 2023,
with a sensitivity around 170 Mpc. The full A+ design sensitivity is expected
only after 2028. We may expect that the LIGO detectors and Virgo detector will
be operating near their A+ design sensitivities of about 260-330 Mpc, and the
KAGRA detector has a goal of reaching above 130 Mpc in late 2020s. The number
of sources in the range goes approximately as the cube of the range (volume).
The scientific relevance of the LIGO-India detector will strongly depend on
its sensitivity in actual operation; \emph{at each stage after commissioning,
it needs to be at a good fraction of the LIGO-US detectors to significantly
contribute to the detection and source localization in the network operation}
\cite{Living-plans}. Due to the cubical dependence of the source-rate on the
strain sensitivity, a detector is relevantly useful in a network during the
mature era of GW observations only if its sensitivity is at least a factor
between 2 to 3 of the LIGO A+ detectors. \emph{Thus, the LIGO-India detector
needs to be sensitive at the level of 100-150 Mpc, with more than 50\% duty
cycle, to be scientifically relevant in the GW advanced detectors network that
is expected to remain operational until the third generation 10 km scale
detectors are deployed and operational, perhaps as early as 2035}.

\section{The Implementation of LIGO-India}

LIGO-India interferometer components are identical to the aLIGO components,
already fabricated for the interferometer previously meant as a third LIGO
detector at the Hanford site, with 2 km arm length. They are in storage in the
US at present. These include both the passive and active vibration isolation
systems, all optical components, electronics (except computers and
peripherals), proprietary instruments for monitoring and characterization etc.
Some of the mirrors need refiguring and re-coating in order to be suitable for
the 4 km LIGO-India detector. Some components of control systems, electronics
hardware and the laser source require technology upgrades. In addition, the
entire package of squeezed light injection and detection is to be fabricated
and set up afresh. If a more pertinent plan that follows the upgrade plans of
the LIGO-USA is adopted, to make the LIGO-India detector consistent with the
\textquotedblleft A+ design\textquotedblright\ (a significant upgrade of
detector components), further infrastructure and instrument modifications
would be necessary \cite{Shoemaker2020}. The design and specifications of the
LIGO-India detector, and the technical tasks for its realization, are clearly
known. There are specialist engineering tasks as well as specialist
commissioning tasks. Both require not only the development of specialist
expertise but also a trained work culture. The need to familiarize with a
rigorously specified and controlled work culture is because this is the first
large scale science project in India with stringent needs of quiet and clean
physical environment as well as the sustained and disciplined monitoring by a
team of experts. The advanced GW detector is a dynamic complex apparatus that
requires refined expertise and familiarity for managing a stable operation.

The trained humanpower required to assemble and commission the LIGO-India
detector is estimated to be about 70-80 members, and about half of them should
be experts in various key areas of physics, engineering, and technology. A few
experts from the LIGO laboratory are expected to participate in some
specialized tasks, especially during the assembly of vital optical elements
and commissioning, but extended direct participation will be limited because
of the ambitious upgrade plans of the operating ground based detectors in the
next decade. This means that the realisation of the operational LIGO-India
detector is crucially dependent on the amount and quality of the expertise
available and trained within the core Indian institutions involved in the project.

\section{LIGO-India Today}

Where is LIGO-India today? The Prime Minister of India announced the support
and notional approval for the project in February 2016. However, various
aspects of the project, including site section and essential characterisation,
have been anticipated and attended to since 2012 onwards. Around the same
time, the KAGRA 3-km underground gravitational wave detector project in Japan
had received full approval, and the construction was in full swing by 2016.
The updated schedule for operation and upgrades of different advanced
detectors, as projected in 2019, is compiled in the figure 1
\cite{Living-plans,Somak2018,Shoemaker2020}. The most recent laboratory that
joined the joint detection program is the Japanese GW detector collaboration
KAGRA. While the LIGO detectors and the Virgo detector are keeping to their
upgrade plans within one or two years, KAGRA has not yet managed to get useful
extragalactic sensitivity. One can see that the schedule projected for the
LIGO-India detector by its management team is lagging already several years
beyond what was originally envisaged. According to the announced schedule, the
detector is supposed to be commissioned for operation in 2026, to join the
network of advanced detectors, with the design sensitivity of about 300 Mpc by
2027. However, an examination of the ground realities paint a completely
different picture, unfortunately.%

\begin{figure}
[ptb]
\begin{center}
\includegraphics[
width=3.5359in
]%
{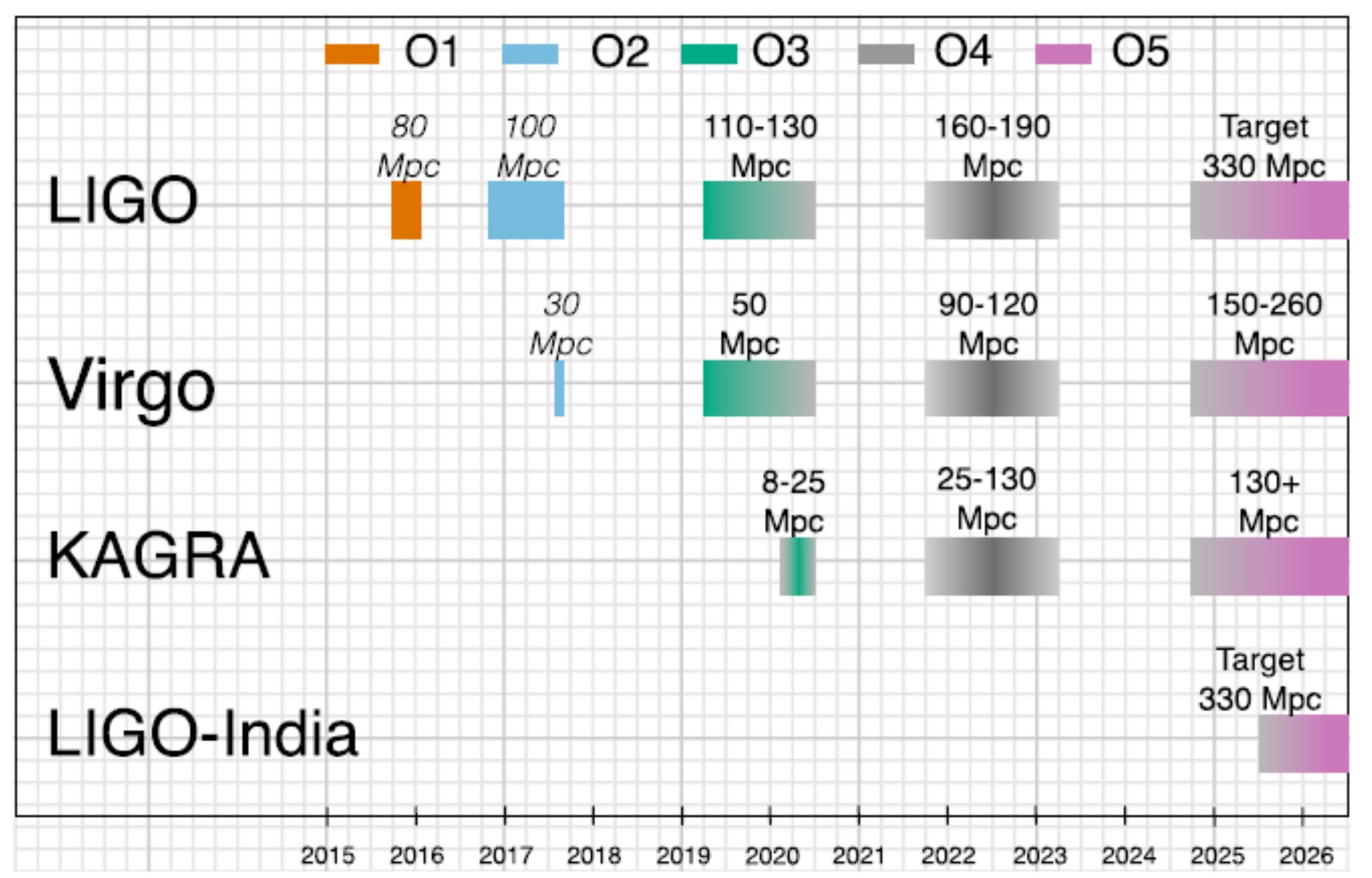}%
\caption{The projected observation schedule as of 2019 for the working LIGO
and Virgo detectors and the commissioned KAGRA detector, along with the
announced schedule for the implementation of the LIGO-India detector
\cite{Shoemaker2020}. This is slightly revised more recently, without a
definite projection for LIGO-India (see fig. 5).}%
\end{center}
\end{figure}

The Nobel prize advanced information document that was released after the 2017
Nobel prizes for the detection of the gravitational waves mentions LIGO-India
(citing my 2013 article on the LIGO-India project and the update in
arXiv:1510.06059) and the Japanese project KAGRA as the future detectors in
the network \cite{Nobel-adv}. KAGRA was scheduled to come online around 2018
and LIGO-India in 2022, as per the original plans. Both projects started
rolling around 2012, though the KAGRA project developed from the large sized
prototypes operated earlier, like the TAMA detector. The project is led by T.
Kajita, who won the Nobel prize in 2015 for the discovery of atmospheric
neutrino oscillations. KAGRA is an advanced detector that incorporated many
new technical features. It uses cryogenic technology for cooling its mirrors
to about 20 K, to suppress thermal noise and thermal lensing. It may benefit
from the \ very low seismic noise in the underground tunnels in the Kamioka
mountain. The tunnel excavation and the fabrication of the UHV\ beam tubes and
chambers were started in 2012 and completed in 2014. It took another 4 years
to complete the assembly of the interferometer and the tests before the start
of commissioning.

The KAGRA interferometer was locked for operation in 2019 and the team is
still working to join network observations, initially with a fraction of the
sensitivity of the aLIGO and Virgo detectors \cite{Kagra-status}. There are
technical difficulties associated with the new design aspects as well as
operational difficulties in maintaining the external environmental conditions
stable. As another example, after decommissioning the initial LIGO
interferometer, an experienced team took about 5 years to assemble and
commission the aLIGO detector with the new components and vibration isolation
systems, using the same infrastructure. A similar time frame was required as
well to bring the upgraded version of the Virgo detector to the level of
network operation.

\emph{The important point to note here is that the examples of operational
detectors demonstrate that it requires 4-5 years to assemble and commission
the detector with an initial working sensitivity after the completion of the
basic infrastructural elements, like the levelled and stabilized land at the
site, the laboratory buildings built according to stringent specifications,
and the fabrication of the beam tubes and the large UHV chambers}. This would
be the case for even an experienced team of about 40-50 scientists, engineers
and technical specialists. The delay is not mainly in fabricating the large
number of components for the interferometer. It is in the careful assembly and
thorough testing at each stage, because even a small compromise in the
operation of individual sections is not an option. Even after achieving the
stable locked operation of the interferometer, reaching a good fraction of the
design sensitivity can take considerable time (typically a year or more),
adding to unpredictable delays. Thus, \emph{even in an optimistic estimate in
the Indian conditions, the minimum duration required for an expert trained
team to realize an advanced interferometric GW detector here with a
network-worthy sensitivity is 10 years from the preparation of the site} -- 4
years for the infrastructure build-up, 4 years for the assembly, tests and
commissioning, and 2 years for tuning the detector for stable locking and
sufficient sensitivity.

The hard lesson is that the time frame for the assembly and initial operation
of the LIGO-India detector will take about 5 years \emph{after the whole
infrastructure of laboratory buildings and the UHV hardware is ready}, if we
manage to have an experienced and skilled team of 50-70 people in the project
by that time, at various levels.
\begin{figure}
[h]
\begin{center}
\includegraphics[
width=3.6197in
]%
{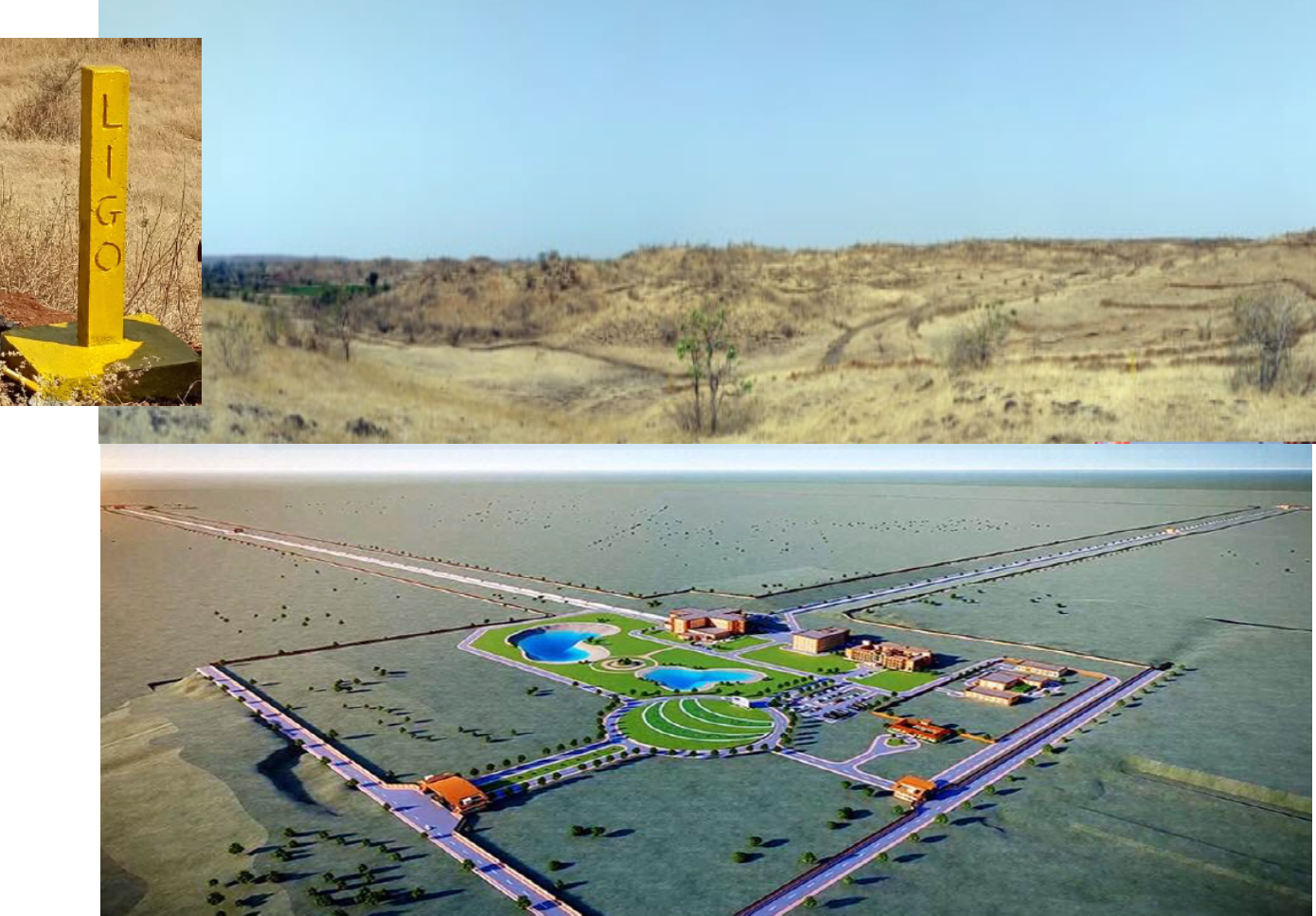}%
\caption{The view of the bare site for the LIGO-India at Aundha near Hingoli,
Maharashtra (2016). The lower panel depicts the vision of the site with the
laboratories and the LIGO-India detector, when the contruction is completed.
The realistic duration to complete this transformation is another 8-10 years,
in my estimate, (see also fig. 4).}%
\label{site}%
\end{center}
\end{figure}

LIGO-India today owns the required land at the chosen site. This is
commendable achievement, made possible by the coordinated efforts of IUCAA,
DAE, government of Maharashtra and the local administration at the district of
Hingoli. Some studies on the soil structure, seismic properties etc. have
progressed. A 16 months baseline seismic survey of the site has been
completed. However, the work on the land preparation for the laboratory
buildings and the levelled beam tube is still in early stages. The chosen site
near Hingoli is accessible by good roads (450 km from Pune, for example).
Flights to the nearby town of Nanded, which already has a small airport, may
become operational. LIGO-India has received the approval for construction from
the state environment impact assessment authority recently. A LIGO-India
office and a visitor guest house have been already set up at Hingoli.

However, the many tasks of building the infrastructure for an aLIGO detector
are yet to acquire any momentum. The huge task of the fabrication of UHV
chambers and the 8 km of beam tube (1.2 m diameter) requires the evaluation
and finalization of the welding technology and strategy, before the
fabrication can be entrusted to industrial partners. This can be a long
process in India. Though the trusted choice for the welding technology is the
advanced version of argon arc welding, the available technology options have
progressed very positively from the days when the LIGO beam tube was built. I
had personally explored new and efficient choices like the K-TIG key-hole
welding, right when I was planning and budgeting for the UHV systems in the
LIGO-India proposal (in 2011). Laser welding is also a technology that has
matured considerably. \ Several tests of both the steel material and the
welding options will be required for a judicious choice that ensures long term
UHV reliability. TIG welding is the chosen method for the fabrication of the
UHV elements of the LIGO-India detector. A large UHV chamber with the
specified design, fabricated in India, is undergoing quality tests. But,
achieving the required level of ultra-high vacuum is still quite far. The
advanced detector requires about a dozen such large chambers. The prototyping
and testing of a beam tube section of 10 m length is yet to be done.

The LIGO-India project and the collaboration do not have a director or a
project manager, even 6 years after the approval by the government of India.
There is a formal management structure for monitoring the project and to
coordinate activities with the LIGO laboratory in the US. \ However, without a
leader, the project has not emerged from its formal confines. Therefore, there
is no close link between the larger GW research community \ and the project
personnel yet. There is a small funding component for small research on next
generation detectors and technology, disbursed to individual researchers in
universities and research institutes other than the LIGO-India core
institutes. But there is no regular interaction possible yet between these
researchers and the LIGO-India project team. As already mentioned, even a
conservative estimate of the expert and trained human power needed to
implement the LIGO-India project is about 70 to 80 participants, for the
different tasks of the assembly of the detector components and vibration
isolation, elaborate position metrology, UHV assembly, leak tests and
certification, optical assembly and internal suspensions, control systems
tests, interferometer locking and full commissioning. \emph{At present only a
fourth of this task force is available}, leading the preparatory tasks in the
core institutes. The LIGO-India project has not yet started a targeted formal
graduate training program. Therefore, there is no significant resource of
young physicists or engineers trained, working with any of the advanced
interferometer detectors, as of now.%

\begin{figure}
[ptb]
\begin{center}
\includegraphics[
width=5.4886in
]%
{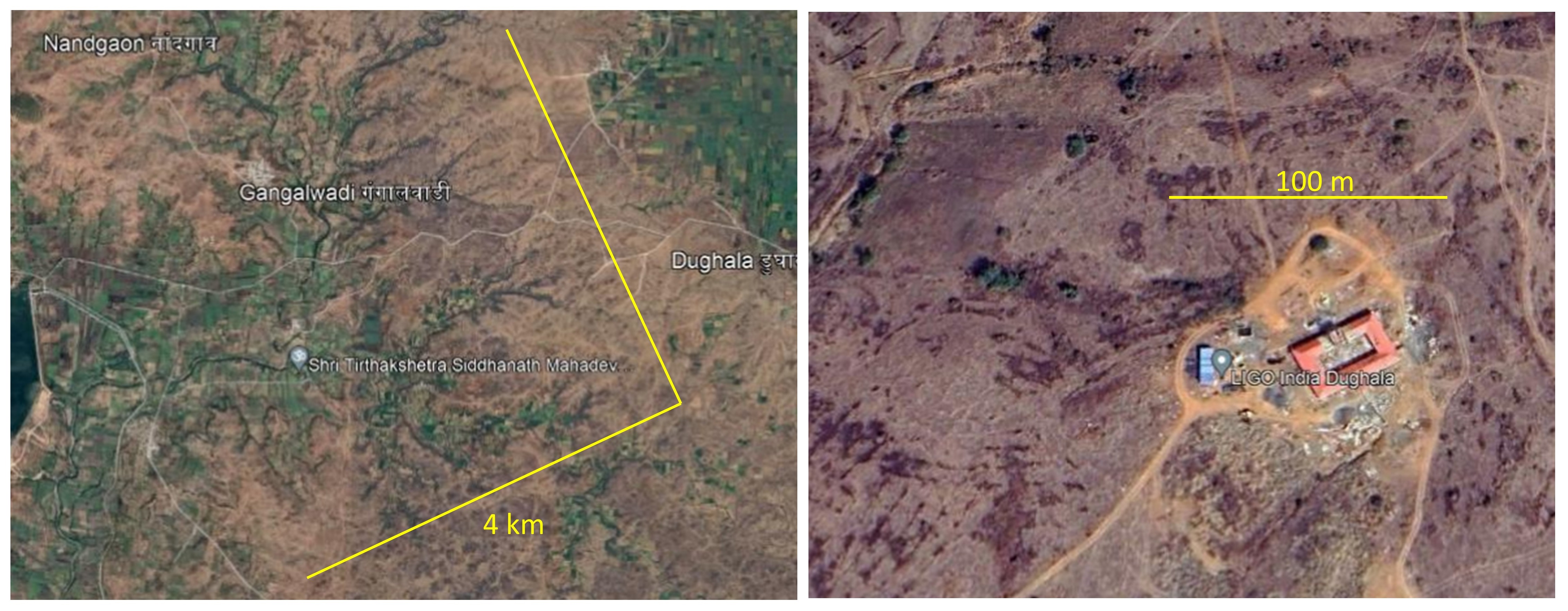}%
\caption{Left) A view of the acquired site meant for the LIGO-India detector.
A sample layout of the 4 km x 4 km detector is indicated. Right) The
construction status as of 2022 November, showing only a site office. The
preparation of the detector site and the construction of the laboratory
buildings are yet to start (\textit{figure extracted from Google Earth}).}%
\end{center}
\end{figure}

The infrastructure of laboratory buildings and the UHV hardware installation
can be readied only after the land is levelled and prepared to the stringent
specifications. The work on major elements of the infrastructure -- the UHV
chambers and beam tubes, site preparation, and the laboratory buildings -- is
yet to start in a scheduled manner. The proper fund flow is perhaps waiting
for an updated detailed project report. This summarises the present status of
the physical progress in the LIGO-India project. Based on the present status
(2022), my experience in the scheduling and budgeting of the original
LIGO-India proposal of 2011 enables me to make a realistic decadal assessment
about the feasibility and scientific relevance of the LIGO-India gravitational
wave detector project. \emph{My revised estimate clearly shows that the
network operation of the detector with a useful extragalactic sensitivity
cannot be earlier than 2032, and the total cost of the project up to 2035 will
be more than doubled, to about Rs. 3500 Crores} (Rs. 35 billion, \$ 430
million). \emph{What is most worrisome is the obvious fact that the large
delay in implementation will severely affect and diminish the scientific
relevance of the project}. I will now examine the various factors that go into
this decadal assessment and revised estimates.

\section{LIGO-India Project Under a `Zeno-Effect'}

The LIGO-India project was about to be launched several years ago, with the
arrow well placed in the bow, still a bit unstretched. Can that arrow reach
its destination when launched? First of all, one needs to wait for the
`launcher' leader(s). Then, as Zeno of Elea wondered, the arrow has to first
reach half the distance before it can progress to the rest. But, it has to
cross `half of half the distance' first, etc. Can the arrow ever leave the bow?

Having studied such situations, and watched many small and big projects in
India and abroad, I know that the arrow can reach the target, if only it is
released. However, in a largely sequential project, the time factor becomes
very important due to the paucity of experienced personnel.
\emph{Gravitational wave detectors work as a network, to achieve the ability
for source localisation and to improve sensitivity and reliability of
detection}. This implies that the detectors in the network ideally should have
comparable sensivities for optimum effectiveness. By the time a delayed
detector reaches a good fraction of its design sensitivity, the already
operational detectors will have progressed to a better sensitivity. Then the
lagging detector will never catch up with the network leaders. Given the
relatively small workforce of limited familiarity with the advanced detector
active in the project, and the many experts who initiated the project expected
to retire half-way through the project, there is a genuine Zeno-effect shadow
on the project. This compromises the relevance and science case of a
gravitational wave detector, meant to operate in a network.

What is an optimistic estimate of the time required to operate the LIGO-India
detector at its A+ design detection sensitivity of about 250-300 Mpc for the
merger of binary neutron stars? What is a reliable road map from initial
commissioning to reaching the design sensitivity and how much time is required?

The preparation of the land, its transformation to the specifications and the
building of the main laboratories will take about 3 years. One need not and
will not wait for these tasks to be completed to start on the enormous task of
fabricating the large UHV chambers and 8 km of the 1.2 m diameter beam tubes.
However, the finalization of the UHV process is yet to be done and the
tendering process is long. Assuming that the fabrication will start in early
2023, we can expect the completed and tested UHV components in late 2025,
ready for assembly. The testing and certification of the UHV hardware is a
long process and it cannot be hurried or shortened. The full infrastructure
(chambers and components in their designated positions in the buildings) can
be expected sometime mid-2026. The assembly and testing of the beam tubes and
their cover etc. can go parallel to the interferometer assembly. The assembly
of the full interferometer will take a minimum of 3 years because of the
complexity and the number of components. Thus, the interferometer could be
ready for locking and commissioning only after mid-2029. Assuming that the
commissioning task will be aided well by some experts from the LIGO
laboratory, a reasonable estimate of locked operation is then 2030. From this
point to achieving a sensitivity at which it can join network operations
(50-100 Mpc for BNS merger) takes between 2-3 years. \emph{Optimistically
assuming a minimum duration for tuning up the sensitivity, the LIGO-India
detector will be network ready only after 2032}. It is clear that the present
projection of 2027 by the LIGO-India management is naive and not feasible. The
optimistic revised date for network operation at extragalactic sensitivity is
2033. However, the projected design sensitivity for an A+ detector (300 Mpc)
can be reached, if at all, only well past 2035.

We have assumed that the presently operating detectors (aVirgo and aLIGO-USA)
will be operating at their design sensitivity of about 250-300 Mpc till 2032.
However, further upgrades are planned for these detectors which need to be
taken into account when we discuss the relevance and science case of
LIGO-India in 2032. Note that even at the sensitivity near 100 Mpc, LIGO-India
will see effectively less than 1 in 25 (%
$<$%
4 \%) of all events detected by the LIGO-US and the Virgo detectors (the
events might be detected, but with much less statistical significance compared
to other detectors). \ Therefore, \emph{though operating as a useful support
instrument for the network, it cannot play a decisive role in discoveries,
unless the sensitivity is increased to the design sensitivity}. However, this
is a long path of tuning and can take many years, by which we enter the era of
next generation detectors. As indicated earlier, in my estimate based on the
progress achieved in presently working advanced detectors, the LIGO-India
detector can hope to reach near its design sensitivity of 250-300 Mpc only in
2035. This should be seen in the context that by 2040, the global detector
network hopes to move up from the present advanced detectors to the next
generation detectors like the Cosmic Explorer and the Einstein Telescope, with
\ 8-10 times the sensitivity of the present advanced detectors.

I am not fully aware of the upgrade plan for the KAGRA detector. But, the
present plan is to be a full partner in GW astronomy mission of the advanced
detectors by joining the next cycle of observation called the `O4' observation
cycle, due in early 2023. This is a crucial time when the additional detector
will help in better localization of the astrophysical sources, even with a
reduced sensitivity, as long as the signal is visible in the detector with a
reasonable signal-to-noise ratio.

The commissioning of an advanced detector in India by 2025-26 was a
scientifically worthy and attractive goal in 2015, when the first indication
of a detection came confirming the wealth of astronomy in store for the
interferometric GW detectors. But, as viewed in 2022, the completion and
operation of a detector after 2032 at the same level of sensitivity (about 100
Mpc) is no more a scientifically exciting goal, though its relevance as an
essential instrument for precision astronomy in a multi-detector network
remains intact. This is because LIGO-India is the only planned detector with a
sensitivity matching the two LIGO instruments in the USA. However, the
significant delay does affect its role as an instrument for discovery.
\emph{What is evident from even an optimistic estimate of the operation
schedule for LIGO-India is that the detector will definitely miss the
opportunity to contribute to multimessenger astronomy during the decade,
2022-2032}. This does not exclude a supportive role for LIGO-India in GW
astronomy in the ensuing decade. However, the enthusiasm in the astronomy
community, even within India, about such a detector will be diminished.

It is very clear that the undesirable situation of the LIGO-India detector
operating always at 3-4 times less sensitivity than other detectors in the
network should be avoided. However, I do not see how this unfortunate
situation could be avoided in any reasonable projection of the schedule,
severely affected by the past delays. This needs urgent and serious discussion
and remedial measures, if there are any.

\section{The Cost Factor}

When the LIGO-India proposal was being prepared in 2011, we were very careful
to minimize costs, and went to great lengths to find out optimal ways to keep
the cost within reasonable demands. What was important in this goal was to
identify expertise within the academic community and institutes, instead of
delegating everything to industrial expertise. In any case, the required
expertise was not available in the industrial circles. Also, timely execution
was a crucial factor. Since the major costs are in UHV fabrication and
certification, it is very important to compare and evaluate the cost increase
in steel and welding technologies etc. From 2015 to 2022, the price of
stainless steel has more than doubled. This alone can increase the costs by
\$80 million. When the proposal was submitted in 2011, the projected budget of
about Rs. 1250 crores (\$270 million then) seemed adequate, with the
understanding that a 15-20\% eventual increase and revision of costs should be
expected by the time it is implemented in 2020. Now in 2022, the projected
costs have more than doubled. A reassessment indicates that project
implementation will require about Rs. 3500 crores (about \$430 million now;
the dollar has significantly increased against the Indian Rupee, by a factor
1.7). At present, the released funding in the past six years is less than 4\%
of the total revised cost.

The Indian government has other megascience commitments as well, both national
and international. The funding required for the international ITER\ project on
nuclear fusion (located in France), which is in construction phase, will be
over 2 billion dollars (%
$>$%
Rs. 15000 crores, estimated as 9\% of 25 billion dollars total international
budget), with the expected commissioning around 2035. The INO project on the
atmospheric neutrino observatory (India-based Neutrino Observatory at Teni,
Tamilnadu, India) was fully approved, and funds were allocated, but it
continues to face legal hurdles sustained by a hostile situation of local
politics and misleading propaganda. The \$300 million project is unlikely to
be realized now because all the members with experience in experimental
particle physics and detector technology have either retired or are about to
retire. The INO graduate training strategy has not resulted in training
adequate expertise among the younger generation to take the project forward.
Thus, there are no experts to lead various aspects of such a project in the
next decade. Also, the science case of the project (if and when it can be
realized) has weakened significantly, due to the delays and the launch of new
projects on neutrino observatories elsewhere. The hyper-Kamiokande project
(with the projected commissioning after 2025) is aimed at solving the
outstanding problems of mass hierarchy, CP violation in neutrino sector etc.
The DUNE project led by the Fermilab also has unique capabilities in answering
such outstanding issues. One can gauge the relevance of the INO detector that
will take 8-10 years for completion, by comparing with these other
initiatives. This makes the INO project scientifically much less significant,
even if all the extraneous hurdles were removed today (which is not the
reality). Then there are commitments in the country in optical and radio
astronomy, for international telescope projects outside the country. The
hugely delayed Thirty-Meter Telescope (TMT) project for the large optical
telescope, proposed to be installed at Hawaii, is another major international
commitment that exceeds Rs. 1500 cores. The (current) commitments in other
astronomy projects like the Square Kilometer Array (SKA) are smaller.
LIGO-India is indeed the largest `home' project in basic science and astronomy
ever taken up in India.

Unfortunately, the revised funding needed for LIGO-India conflicts with the
present situation of decreased GDP growth in India and elsewhere. The Covid-19
pandemic is a global problem, but it will seriously affect and delay the fund
flow into the mega-science projects, justifiably. No substantial quantum of
funds is approved or released yet for LIGO-India because the actual spending
for the large hardware and infrastructure has not yet started. In my
judgement, the cost factor will significantly impact the LIGO-India schedule,
at least in the early part of its execution, and will unpredictably delay the
operation of the detector. \emph{On the other hand, any decision on fully
funding the project now will have to evaluate and reconsider the \ doubled
cost factor for a project that has reduced scientific significance, when
implemented for operation after 2032}.

\section{A View from the Other Side}

The LIGO-India project is not merely an Indian national project. It is an
important multi-national collaborative effort, of setting up and operating one
of three identical advanced LIGO detectors in India, in the network operation
with the other two detectors in the USA, and also with the Virgo detector in
Europe. An MoU was signed with the National Science Foundation and the LIGO
laboratories (Caltech and MIT). It cannot be that the NSF and LIGO
laboratories are unaware of the steadily diminishing scientific and
astronomical relevance of the delayed LIGO-India project. In fact, the
executive director of the LIGO laboratory, David Reitze, had noted in a
meeting in India (Pune, 2017) that \emph{a serious risk to LIGO-India
scientific program exists if delays begin to accumulate}.

The commitment by the LIGO laboratory for providing all the interferometer
components, some suitably modified, still stands. These items that are stored
in safe packing await shipping to India. However, before the shipping to India
takes place, the infrastructure to receive them needs to be completely ready.
In the absence of timely progress on this front, what are the options at this stage?

Apart from the uncertain internal cost factor involved in the commitment of
providing the interferometer components, there are even changes in the basic
rules of the import of scientific equipment to India. Unless explicitly
exempted, there is an additional burden of at least 18\% cost as Goods and
Services Tax, to be collected as additional customs duty, for equipment
imported by scientific establishments. Of course, this will have to be
provided as additional funding to the LIGO-India project by the Government of
India itself -- sort of a strange situation of giving with one hand and taking
with another. But, the added administrative cost on both sides is not negligible.

Even at a reduced scientific relevance, the only possibility for gainfully
using the components fabricated for a third LIGO interferometer seems to be
the implementation of the LIGO-India project. In my view, there seems to be no
possibility of reviving a plan like LIGO-Australia. However, a proper
evaluation of the projected joint operation of the ground based detectors,
estimating accurately the efficacy of the LIGO-India detector in network
operation is necessary. It is clear that the LIGO-India detector will remain
in a supporting role unless a straggling gap in the projected sensitivity,
compared with other operational detectors, is mitigated. I will now examine a
path for regaining a prominant role for LIGO-India detector.

\section{LIGO-India A+}%

\begin{figure}
[ptb]
\begin{center}
\includegraphics[
width=3.3948in
]%
{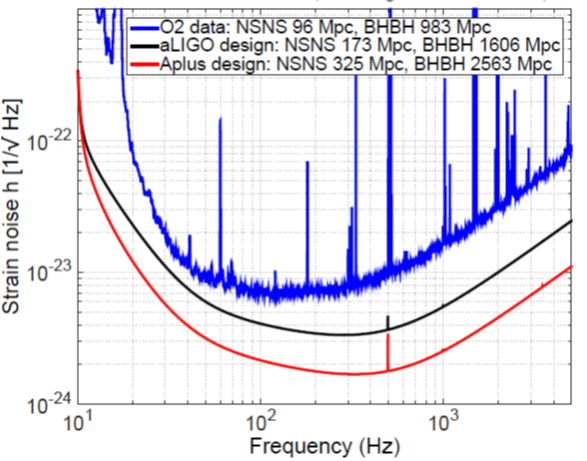}%
\caption{Projected sensitivity of the A+ upgrade in 2026, compared to the
sensitivity of presently operating aLIGO \ detectors
\cite{Shoemaker2020,McCuller2019}. }%
\label{A+}%
\end{center}
\end{figure}

There is a well directed plan for a significant upgrade of aLIGO detectors to
LIGO A+ detector, to be completed by 2026 \cite{Shoemaker2020,McCuller2019}.
The technology additions envisaged are mature squeezed light quantum
measurement and vastly improved low loss mirror coatings. There are other
upgrade options as well, by adding to the interferometer baseline. Clearly,
LIGO-India should directly be installed and commissioned in its A+ version,
since its assembly phase is definitely after 2025. If this is not done
LIGO-India will not be very relevant or useful in the network, when it is
fully commissioned. Therefore, \emph{the present understanding is that
LIGO-India will be installed in its A+ version directly}. Still, LIGO-India
needs to go through the long and elaborate process of assembly, commissioning,
and step by step tuning of the sensitivity to reach the sensitivity suitable
for the global network operation. Since it is clear that the LIGO-India
detector is behind its announced schedule by at least 5 years, it is now
excluded from consideration for the quantitative estimates for the efficacy
(localization of sources, network duty cycle etc.) in the parameter estimation
from the observations during 2026-2030.

The O5 observation cycle for the LIGO-Virgo detectors is planned during
2026-2029. The LIGO detectors will operate with A+ design during O5. However,
the post-O5 operation of the LIGO detectors envisages a major upgrade called
\textquotedblleft A$^{\#}$\textquotedblright\ (written as A\# in the rest of
this paper) \cite{LIGO-Ahash}. This is expected to operate with two times
better sensitivity than A+. \emph{I am convinced that this situation demands
an immediate redefinition of the plans for LIGO-India, because it will be
ready for observations only after the LIGO-A\# detectors are commissioned in
the USA for their post-O5 operation}. I discuss this vital point next.%

\begin{figure}
[ptb]
\begin{center}
\includegraphics[
width=5.5973in
]%
{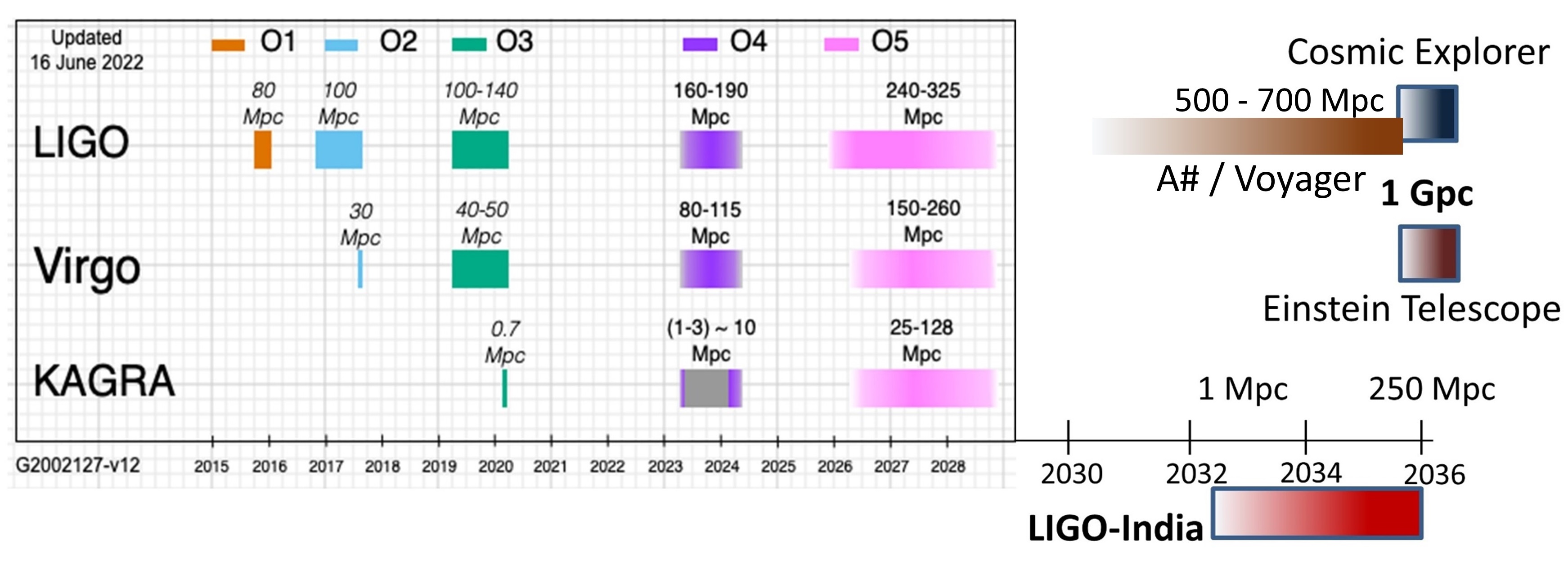}%
\caption{Revised schedule for the observation cycles with progresively
increasing sensitivity of LIGO,Virgo, and KAGRA detectors, post-pandemic (LIGO
document G2002127). I have added a realistic estimate for LIGO-India. By the
time the LIGO-India detector is operational, gravitational wave astronomy will
have entered an era of next generation detectors, like the Voyager upgrade in
the LIGO facility. Post 2036, we can expect the beginnings of the 10 km class
Cosmic Explorer and Einstein Telescope.}%
\end{center}
\end{figure}

\section{Beyond A+: Post-O5 and LIGO-A\#}

Further improvement of the LIGO-USA facility beyond 2029, after the
observation cycle O5, have been in discussion for some time, and two
possibilities were studied. The option called \textquotedblleft
A\#\textquotedblright\ involves replacing the main test mass mirrors with much
heavier mirrors and suitable upgraded suspension and vibration isolation
systems. The A+ version has 40 kg mirrors whereas the A\# design is with 100
kg test masses, suspended on the silica fibres at twice the stress levels.
Also, the intra-cavity laser power at the same wavelength (1064 nm) will be
increased from 750 kW to 1500 kW. This requires better reflective coatings of
the mirrors, still under development. The thermal noise in the mirror coating
is expected to be reduced by a\ factor of 2. Further, the amount of quantum
squeezing of light at the detection section will be improved (from 6dB in A+)
to 10 dB. These are the main upgrades that will result in the superior
sensitivity of the A\# design.

The A\# option is at present preferred over the second option called
\textquotedblleft Voyager\textquotedblright, in which the existing optical
elements are to be replaced with entirely new optics made of Silicon. The
suspended Si mirrors will be much heavier, about 200 kg. These will be in a
semi-cryogenic environment of 123 K, at which the low mechanical loss and the
zero thermal expansion of Silicon will result in reduced thermal noise and
thermal lensing. The same UHV infrastructure would be used for this transition
to the Voyager version \cite{Voyager}. The high power stabilized laser will be
replaced with one of longer wavelength, near 2 micron, to match the reflection
characteristics of the Silicon optics. The projected improvement in
sensitivity of these options is by more than a factor of 2, or about 700 Mpc
for BNS merger.

There has been no serious discussion so far about the role of LIGO-India in
its A+ version during the era of A\# or Voyager. If the technology tests
succeed and the A\# version of the LIGO-US detectors is implemented in early
2030s, LIGO-India will only have a fraction of the sensitivity of the LIGO-US
detectors, when it becomes eventually operational in 2032. Therefore, a
definite road map to closely participate in the LIGO developments and to
follow its course needs to be framed soon. In other words, the post-O5 plan of
the A\# version (or Voyager) of LIGO-USA detectors makes it imperative that
LIGO-India follows the same course, for operation at a good fraction of the
sensitivity of the leading LIGO detectors. \emph{Aiming to implement the lower
sensitivity A+ version for LIGO-India is certainly not the right course,
according to my estimate and projections, if we desire a frontline stature in
the GW detector network.} However, building the detector with the A\# design
requires certain incremental changes in the main infrastructure (vibration
isolation, UHV chambers, layout etc.) \emph{as well as the additional funding
required for the upgraded elements}.

\section{Can LIGO-India be a Late Yet Significant Success?}

This question has a conditional affirmative answer, which I examine now. What
is definite about the LIGO-India project is that it will miss the O5
observation phase of the LIGO-Virgo-KAGRA detectors, scheduled from late 2026
for a few years. We need to look for a significant participation beyond O5.
The analysis so far, especially the aspect about a reduced scientific
relevance, was based on the announced projections on the systematic upgrades
and developments beyond LIGO A+ design, which is the one to be implemented in
LIGO-India. However, these projections naturally have some uncertainty arising
from technical difficulties of cutting edge technology, as well as from
external factors like funding, global social and economic situation etc., as
one can understand. A very favourable aspect that guaranteed a special
position for LIGO-India in the detector network was the fact that its design
is identical to the two LIGO detectors in the USA, which operate at a higher
sensitivity than advanced Virgo, given it smaller baseline of 3 km. The
KAGRA\ detector at its design sensitivity cannot match with even Virgo. Thus,
the possibility of a late LIGO-India to be the crucial element in achieving
localization capability of the detector network at the level of a few square
degrees is real and evident, especially if KAGRA fails to achieve the promised
design sensitivity beyond 130 Mpc even by 2032. The fact that LIGO-India will
replicate the successful design elements of the LIGO-USA detectors is its
reliable advantage. However, the duration to reach a decisive sensitivity
cannot be reduced significantly from the estimate that I discussed already.
Given this situation, if LIGO-India is commissioned with an initial
extragalactic sensitivity larger than 10 Mpc, before 2032, and tuned up to a
stable sensitivity of 300 Mpc by 2034, then it will enjoy a position of
prominence with the LIGO-Virgo network for a few years, till \ about 2040,
with progressive improvement to the design sensitivity beyond 500 Mpc.
However,\emph{ it is mandatory that LIGO-India is replanned with the A\#
design meant for post-O5 era, rather than the less sensitive A+ design, for it
to be prominant and truly relevant}. Such a major shift of focus is what I
propose and advocate, because \emph{I think that the A\# design is the right
option for a delayed LIGO-India}. This is certainly feasible by revamping the
present structure of management and work culture, along with a determined
action on all aspects of human resource development. However, this feat is
crucially dependent on the availability of necessary funds and significantly
augmented expert human resources, after the full approval of a revised DPR and
the project within the next two years. It is a small window of opportunity.
Again, the absence of a leading figure of science and technology to take the
LIGO-India project forward at the required pace is clearly felt. The five
factors -- reliable leadership, expert human power, funding, time factor, and
scientific relevance -- need to be urgently addressed and reviewed carefully
for any realistic strategy of execution.\footnote{I am a member of the LIGO
Scientific Collaboration (LSC) and the LIGO-India Scientific Collaboration
(LISC). This article is based entirely on my presonal perception and estimates
about the LIGO-India project.}

\section{Summary}

In my earlier overview paper (2013) on the scope and plans for the LIGO-India
gravitational wave detector, I sketched some key developments on the large
canvas of the megascince project. It was then perceived as transformational
for Indian science, technology and higher education, and rejuvenating for
Indian physics and astronomy. More than a decade after the LIGO-India
proposal, and six years after its notional approval, the actual construction
of the infrastructure and the fabrication of the critical detector elements
are yet to start. This implies a shift in the schedule for implementation by
several years, and the integration in the global network by a decade, to
beyond 2032. Also, there are other hindering factors, like substantial cost
escalation, retirement of current experts etc. The consequences for the
science scope of the project and the definite erosion of the relevance of the
delayed LIGO-India detector in a network operation, in the era of GW and
multimessenger astronomy, are examined. The LIGO-India detector is sure to
miss the seminal decades of multi-messenger astronomy, 2015-2035, when various
kinds of sources are being discovered and many fundamental aspects are being
tested. However, \emph{with timely revamped action, the LIGO-India project
with a A\# design can still be a late yet significant success}, serving a
prominent role in the detector network and precision GW astronomy for a decade
after 2034. My assessment urges \ an immediate evaluation and restructuring of
the LIGO-India project in the context of the unavoidable stretch in its schedule.

\section*{Acknowledgements}

This article benefitted immensely by the suggestions and comments by David
Shoemaker. I thank Martine Armand for many helpful comments.

\end{document}